\documentclass[pdflatex,sn-mathphys-num]{sn-jnl}

\usepackage{graphicx}%
\usepackage{multirow}%
\usepackage{amsmath,amssymb,amsfonts}%
\usepackage{amsthm}%
\usepackage{mathrsfs}%
\usepackage[title]{appendix}%
\usepackage{xcolor}%
\usepackage{textcomp}%
\usepackage{manyfoot}%
\usepackage{booktabs}%
\usepackage{algorithm}%
\usepackage{algorithmicx}%
\usepackage{algpseudocode}%
\usepackage{listings}%
\usepackage{tcolorbox}
\usepackage{float}

\usepackage{graphicx}
\usepackage{natbib}
\usepackage{doi}
\usepackage{caption}
\usepackage{array}

\theoremstyle{thmstyleone}%
\theoremstyle{thmstyletwo}%

\theoremstyle{thmstylethree}%

\begin{document}


\title{Prioritizing Software Requirements Using Large Language Models}

\author*[1]{\fnm{Malik Abdul} \sur{Sami}}\email{malik.sami@tuni.fi}
\author*[1]{ \fnm{Zeeshan} \sur{Rasheed} } \email{zeeshan.rasheed@tuni.fi}
\author*[2]{\fnm{Muhammad} \sur{Waseem}}\email{muhammad.m.waseem@jyu.fi}

\author[]{\fnm{Zheying} \sur{Zhang}}\email{zheying.zhang@tuni.fi}

\author[]{\fnm{Tomas} \sur{Herda}}\email{herda.tom@gmail.com}

\author[]{\fnm{Pekka} \sur{Abrahamsson}}\email{pekka.abrahamsson@tuni.fi}

\affil[1]{\orgname{Tampere University}, \orgaddress{\city{Tampere}, \country{Finland}}}
\affil*[2]{\orgdiv{Faculty of Information Technology}, \orgname{University of Jyväskylä}, \orgaddress{\city{Jyväskylä}, \country{Finland}}}









\abstract{Large Language Models (LLMs) are revolutionizing Software Engineering (SE) by introducing innovative methods for tasks such as collecting requirements, designing software, generating code, and creating test cases, among others. This article focuses on requirements engineering, typically seen as the initial phase of software development that involves multiple system stakeholders. Despite its key role, the challenge of identifying requirements and satisfying all stakeholders within time and budget constraints remains significant. To address the challenges in requirements engineering, this study introduces a web-based software tool utilizing AI agents and prompt engineering to automate task prioritization and apply diverse prioritization techniques, aimed at enhancing project management within the agile framework. This approach seeks to transform the prioritization of agile requirements, tackling the substantial challenge of meeting stakeholder needs within set time and budget limits. Furthermore, the source code of our developed prototype is available on GitHub, allowing for further experimentation and prioritization of requirements, facilitating research and practical application.} 

\keywords{Software Engineering, Requirements Engineering, Requirements prioritization, LLMs, Generative AI }



\maketitle

\section{Introduction}\label{sec1}

Agile project management is a prominent methodology in software development, known for its flexibility and responsiveness to continuous changes \cite{lei2024artificial}. Some leading Agile methodologies include Extreme Programming (XP), Scrum, and Kanban.\cite{9487986}. Each of these methodologies has unique practices and principles designed to address specific aspects of project management and software development; for instance: fixed-length iterations and defined roles, such as the Scrum Master and Product Owner, are key components of Scrum. Kanban focuses on improving workflow through the visualization of work and the limitation of tasks. XP aims to enhance product quality by emphasizing frequent releases, simplicity, and continuous feedback \cite{abrahamsson2017agile}. In addition to these approaches, more sophisticated, global, or corporate contexts are adopting scaled agile frameworks like SAFe (Scaled Agile Framework), LeSS (Large Scale Scrum) \cite{carroll2023transformation, saklamaeva2023potential}. These frameworks take on the issues of scalability and ensure that Agile approaches are successfully used in bigger projects and organizations by extending Agile principles to coordinate and align different teams. However, In response to the imperatives of agile methodologies and the increasing demands for rapid development, the field of software engineering is evolving at an accelerated pace, therefore it's critical to identify effective, accurate, and adaptable methods for prioritizing Software requirements \cite{svensson2024not}. 

Artificial Intelligence (AI) and NLP have ushered in a new era of computational tools capable of understanding and generating text that closely mimics human writing with remarkable accuracy. The development of powerful language models, particularly those in the GPT series, marks a significant milestone in this journey \cite{radford2018improving,rasheed2024can, ouyang2022training}. These LLMs, with their advanced natural language understanding capabilities, stand at the forefront of a paradigm shift in software engineering. They transition from being mere tools to acting as collaborators that offer insights and analysis previously unattainable through conventional methods \cite{sami2024system,  rasheed2023autonomous}. The integration of LLMs into the requirements prioritization process is not merely an academic exploration but a necessary evolution in the dynamic landscape of software development, addressing the urgent need to automate, enhance, and refine how requirements are prioritized \cite{tikayat2023agile, zhang2024llm}.

Despite the widespread adoption of agile methodologies in software development, teams continue to struggle with effectively prioritizing project requirements. This challenge arises from the subjective nature of traditional prioritization techniques, the inherent complexity of evolving project scopes, and the difficulty of swiftly adapting to stakeholder feedback and market trends \cite{tosic2023artificial}. The integration of LLM-based agents in the requirements prioritization process offers a promising solution by automating the evaluation and prioritization of software requirements. However, the practical application of these technologies raises questions about their effectiveness, adaptability, and the balance between automation and human oversight in agile project management.

This study explores the potential of LLMs to improve requirements prioritization within Agile methodologies, focusing on their application and impact on software requirement engineering processes. Our investigation revolves around the following key questions:

\begin{enumerate}
    \item How can large language models be applied in prioritizing requirements?
    \item What are the limitations and opportunities of using large language models for requirements prioritization?
\end{enumerate}

Our research strategy involves using LLMs to:
\begin{enumerate}
    \item Develop a tool that captures requirements from users and generates user stories.
    \item Implement prioritization logic to evaluate and prioritize user stories based on business value, urgency, and technical complexity.
    \item Ensure seamless integration with project management tools like JIRA, Trello, and Azure DevOps to enhance the Agile process and stakeholder coordination.
\end{enumerate}
 
In this paper, we present a web-based software tool for software requirement prioritization, showcasing how GPT models could redefine software engineering. Our prototype illustrates the potential of GPT agents to transform user narrative creation and requirement prioritization, empowering stakeholders and fostering collaboration. By accelerating project timelines and addressing trust, time, and budget constraints more effectively, this approach sets a new standard for stakeholder engagement and software development processes with the help of GPT models.

The paper is structured as follows: Section 2 provides a background and a survey of related literature. Section 3 describes the proposed research design and methodology. Section 4 discusses the preliminary results. Section 5 outlines limitations and future work. Finally, Section 6 concludes the paper.

\section{Background And Related Work}
\label{Background}

\subsection{\textbf{Background:} Generative AI and Its Application in Software Requirements Prioritization}

\label{Generative AI}

Generative AI, aims to produce new material by mimicking human-like outputs in a variety of fields, including computer vision, NLP, and the creation of images and videos \cite{hacker2023regulating}. Text creation, machine translation, dialog systems, and code generation have all been transformed by this technology, especially through the use of autoregressive language models such as GPT and Generative Adversarial Networks (GANs) \cite{aydin2023chatgpt}. The transformer architecture, which greatly improves NLP skills by capturing contextual relationships inside text, is the foundation of the GPT model, which is well-known for its human-like text production \cite{radford2018improving,rasheed2024large}.

The exceptional efficiency and adaptability of GPT models, as evidenced by recent advances, indicate that they have tremendous potential to revolutionize SE processes \cite{liu2023gpt,ouyang2022training}. These models can streamline the software development lifecycle by automating tasks like error detection, code snippet generation, and documentation creation after being trained on large-scale code repositories \cite{feng2023investigating,treude2023navigating}. In addition to speeding up code generation and program development, the incorporation of GPT models into SE has improved software development's overall quality and efficiency \cite{dong2023self,ma2023scope}.

Software engineering (SE) has witnessed a rapid evolution of generative AI, demonstrating its transformational power. A major change that promises to accelerate and improve software development is the integration of GPT models into SE. These approaches transform not just the software development process but also the way software is developed and maintained. Most importantly, they improve the prioritizing of requirements, which is an essential development stage. With the use of AI, developers can more efficiently prioritize and order work, making sure that important features match user requirements and organizational goals. The development and maintenance of software will be drastically changed by this method, which represents a shift towards more agile, user-focused, and efficient software development.

\subsection{Related Work}

An approach to needs prioritization in agile software development that is gamified is examined in the paper "PRIUS: Applying Gamification to User Stories Prioritization". To increase stakeholder participation, it blends gaming aspects with the Wiegers Matrix prioritizing technique \cite{lencastre2024prius}. Positive effects on engagement levels during requirements prioritizing were observed in an academic setting during the development and testing of a system that supports this technique.

Several studies (e.g., \cite{ahmad2023towards}, \cite{barke2023grounded},\cite{tikayat2023agile}) have explored GPT models in SE and in requirements prioritization techniques. The application of Large Language Models (LLMs) to software development, specifically for needs prioritization, is a significant advancement that holds the potential to increase both efficiency and effectiveness. However, the problems with hallucinations and the constraints of natural language identify an important research need. A solution that can not only understand broad client needs and create user stories on its own but also uses AI agents to systematically prioritize tasks is desperately needed. The latest software development approaches that emphasize agility and the creative use of LLM agents must be embraced by this instrument. By closing this gap, we hope to improve needs prioritization accuracy and dependability while also fully utilizing AI's capabilities within agile software engineering frameworks. This presents a substantial opportunity for innovative research and development in the sector.

Some established techniques and methods prioritize requirements in software development and project management. Understanding these methods matters for effective decision-making and resource allocation in enterprise-level software projects. While there are other notable techniques, we focus on those we use in our prioritization process. The Analytic Hierarchy Process (AHP) stands out. It blends mathematics and psychology, using pairwise comparisons and expert judgments to create priority scales. This process is known for its mathematical rigor and its ability to mix subjective and objective decision-making elements. Similarly, the MoSCoW method offers a straightforward yet effective framework for prioritizing requirements by sorting them into four categories: Must have, Should have, Could have, and Won't have. This method helps clarify the importance and urgency of each requirement, supporting more informed decision-making.







\section{Preliminary Methodological Framework}

We utilize a web-based tool in our research, employing LLMs to offer a user-centric solution for the precise and efficient prioritization of software development requirements. This tool is enhanced through the integration of React, Flask, and OpenAI technologies. 
For a detailed illustration, we present our methodology steps and their application, providing a visual guide to our approach, the tool’s capabilities, and its output analysis.

\begin{enumerate}
    \item \textbf{Tool Development:} The primary step involves building a web application that uses React for an intuitive front end, Flask for the back end, and OpenAI's APIs to leverage LLMs. This application is designed to automatically generate user stories from input requirements or documents uploaded by users. It demonstrates the practical use of LLMs in creating and prioritizing software requirements according to client needs.
    
    \item \textbf{Requirement Input and Prioritization:} Users can input their requirements or upload predefined user stories. The system uses LLMs to generate user stories and assigns them to epics randomly. Users can choose from various prioritization methods, such as AHP and MoSCoW, for a customized prioritization process, and the system displays the results.
    
    \item \textbf{Case Study Implementation:} We conduct case studies by inputting user requirements of a Systematic Literature Review (SLR) to test the tool's effectiveness in generating and prioritizing user stories. The tool's capability to process and understand requirements in natural language enhances the practical evaluation, ensuring a user-friendly experience and aligning outputs closely with project needs. The prioritized list of requirements and user stories can be downloaded in CSV format for further analysis and documentation.
    
    \item \textbf{Evaluation and Output:} During the evaluation phase, we analyze the tool's outputs, including performance analysis, content analysis generated by LLMs, and stakeholder satisfaction through qualitative analysis.
\end{enumerate}

Our methodology not only empirically validates the effectiveness of our tool but also highlights the practical advantages of integrating LLMs into the requirements engineering lifecycle. It significantly improves the efficiency and alignment of requirements prioritization with modern software development practices.


\begin{figure*}[t]
    \centering
    \includegraphics[width=\linewidth]{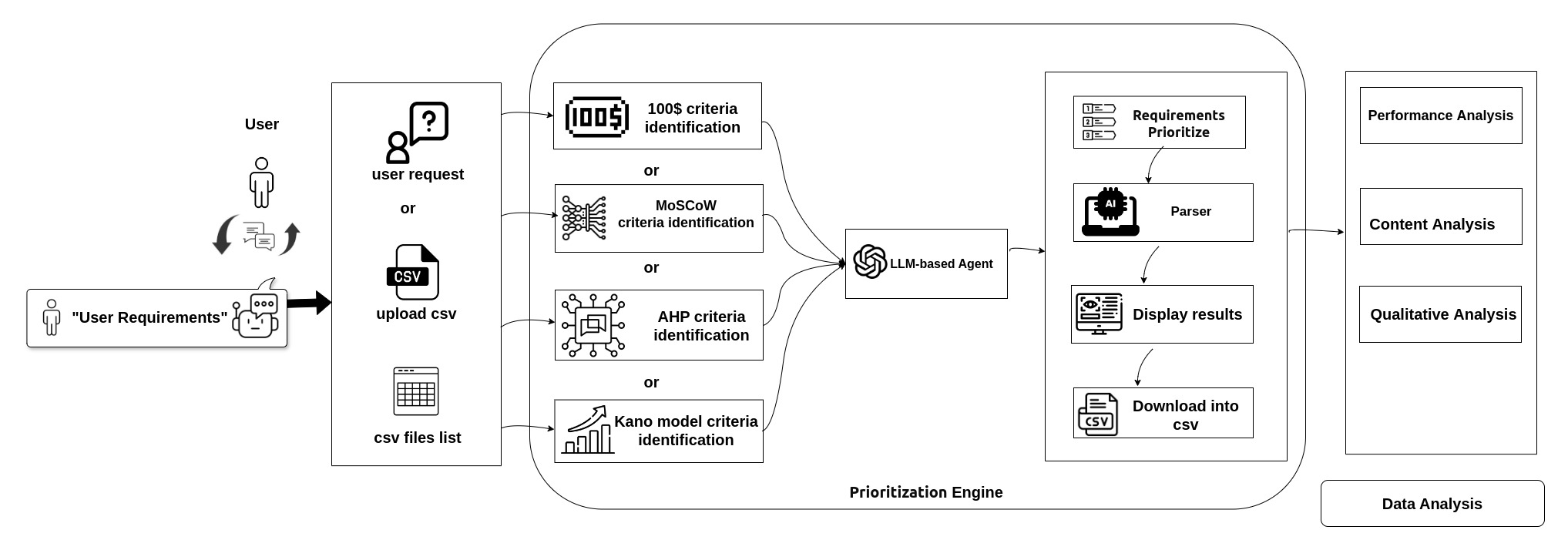}
    \captionsetup{justification=centering}
    \caption{Process of Prioritizing Requirements Using LLMs and Analyzing Its Output}
    \label{fig:result}
\end{figure*}

\section{Preliminary Emperical Results}
\label{preliminary result}

This research aims to increase the effectiveness of research procedures by introducing a system that simplifies the process of creating and ranking user stories. The system seeks to improve research workflows with the use of Python Flask API, React UI, and ChatGPT 3.5. We focus our first assessment on the creation of user stories, how they are prioritized using different methods like MoSCoW, the 100 Dollar Test, and the Analytic Hierarchy Process (AHP), as well as how the user interface is evaluated. We are ready for more investigation after this preliminary analysis. It is anticipated that a future case study, set in a real research setting, will provide hard proof of the system's efficacy and identify areas for improvement based on user input.

\subsection{Theme and Objective}
This case study presents a system engineered to automate the Systematic Literature Review (SLR) process utilizing Language Model (LLM) technologies. Its principal aim is to streamline the generation and prioritization of user stories from delineated requirements, thereby enhancing the SLR workflow's efficiency.

\subsection{Identified User Requirements}
Developed in direct response to specific user requirements, the system is designed to significantly improve research workflows. These requirements include the automated generation of research-specific questions, an intuitive user interface for seamless user-system interaction, flexibility in language model selection, efficient literature review management adhering to predefined criteria, and support for generating crucial research documentation. Meeting these requirements is essential for optimizing research output and process efficiency.

\subsection{System Configuration}
The system's architecture employs React, CSS, and JavaScript with ANT design for the front end, alongside Flask for back-end development. This configuration enables an API that interfaces with the GPT-3.5 model, facilitating the transformation of user inputs into actionable user stories through advanced natural language processing and rendering them back to the user in JSON format.

\subsection{Step-by-Step Operation}

\textbf{Step 1: Understanding user needs} The system prompts users to input their initial requirements and desired number of user stories through an engaging interface. It then processes these inputs, parses the results, and displays them promptly, as depicted in Fig 2.

\begin{figure*}[t]
    \centering
     \includegraphics[width=\linewidth]{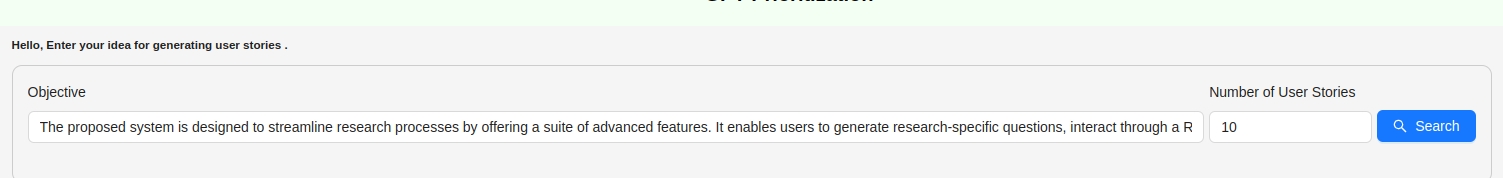}
    \captionsetup{justification=centering}
    \caption{Requirements Gathering}
    \label{fig:result}
\end{figure*}

\textbf{Step 2: Displaying results} The parsed data, presented in JSON format, is displayed within a dynamically generated React table (ANT design), offering a detailed view of the user stories, as shown in Figure 3.

\begin{figure*}[t]
    \centering
    \includegraphics[width=\linewidth]{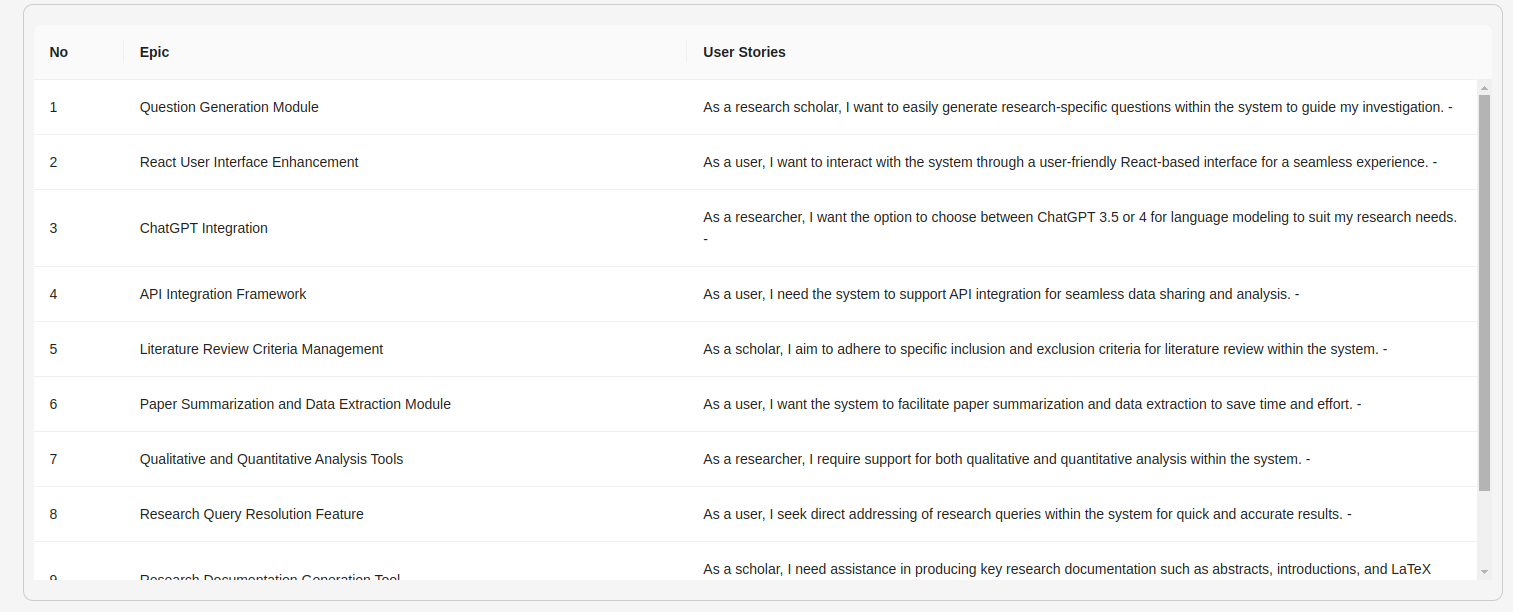}
    \captionsetup{justification=centering}
    \caption{Generation of User Stories by LLMs}
    \label{fig:result}
\end{figure*}

\textbf{Step 3: Prioritizing Requirements} Subsequently, the system provides a feature for prioritizing these requirements using various techniques, including the Analytic Hierarchy Process (AHP). Users input their prioritization criteria, and the system organizes the user stories accordingly, converting the prioritized list back into JSON format. This prioritization process is illustrated in Figure 4.

\begin{figure*}[t]
    
     \includegraphics[width=\linewidth]{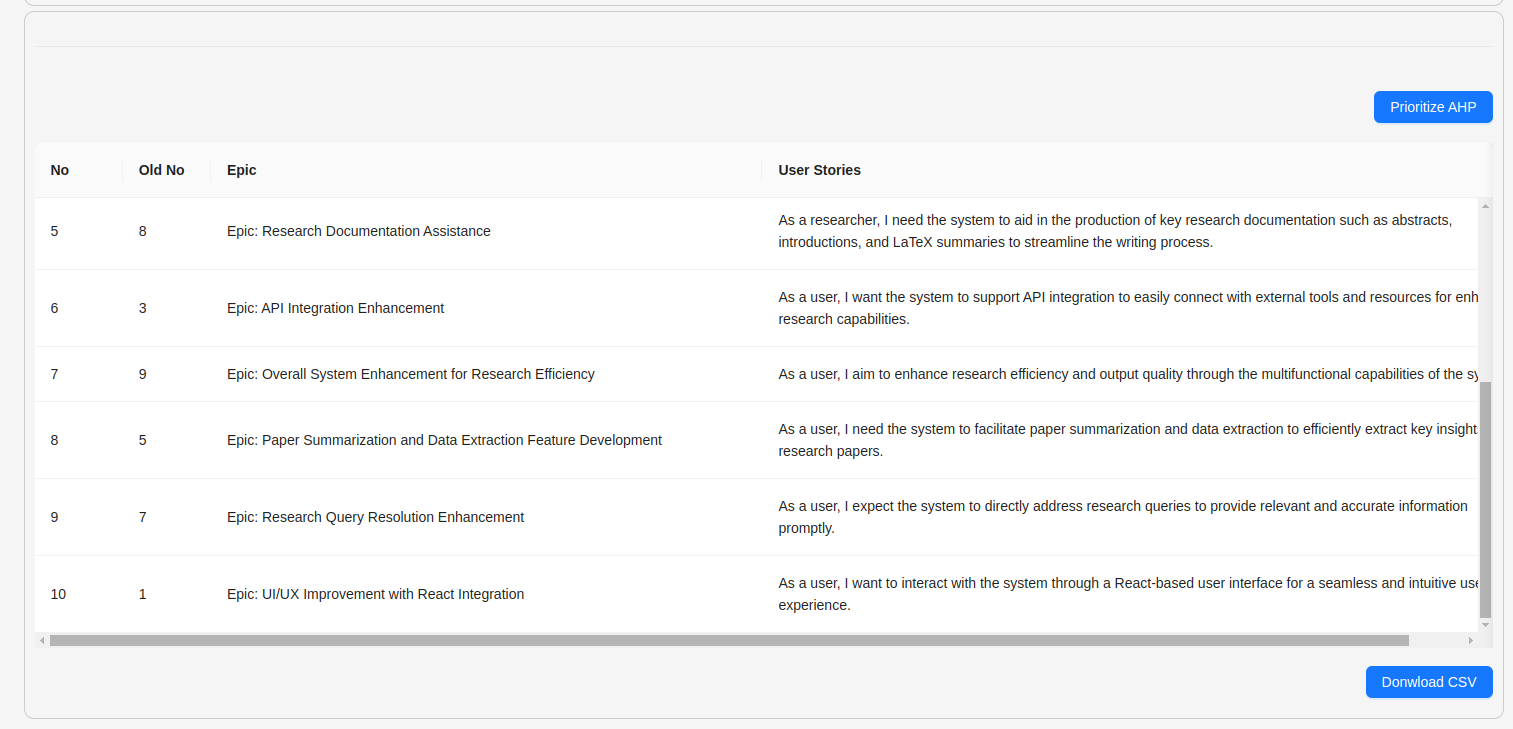}
    \captionsetup{justification=centering}
    \caption{Prioritization of Requirements Using LLMs}
    \label{fig:result}
\end{figure*}

\textbf{Step 4: Exporting Requirements:} Ultimately, the system makes it easy to export prioritized requirements in CSV format. This allows for simple integration into project management tools such as JIRA, Azure DevOps, and RE management tools.

\subsection{AHP Analysis and Case Study Performance}

This section explores the system's capacity to automate essential aspects of the SLR process effectively, particularly highlighting its proficiency in generating and prioritizing user stories based on specific user needs through the Analytic Hierarchy Process (AHP). Our approach uses AHP for user story prioritization, focusing on three main criteria: Business Value, Technical Feasibility, and User Impact, which collectively ensure that development efforts align with business objectives, maintain technical integrity, and significantly improve user experiences.


Efficiency is demonstrated by the AHP-based prioritization process completed in an average of eight seconds, showcasing the method's effectiveness and the system's capability to quicken the user story generation and prioritization phases. Moreover, initial observations reveal the system's efficiency in generating epics and user stories from a minimal set of input requirements, significantly outperforming manual methodologies.

\section{Limitations and Future work}
\label{limitations}
Although this study uses Large Language Models (LLMs) to improve requirements prioritization within agile approaches, it acknowledges several limitations that open up new ways for future research. Mostly, LLM hallucinations pose a difficulty; these models tend to produce outputs that don't match the input data \cite{xu2024hallucination}. So we have to undertake a amount of testing and validation to make sure the generated user stories are reliable. Though limited in scale, we provide preliminary proof of concept that highlights the potential of LLMs in simplifying the requirements prioritizing process through the creation of basic user stories from requirements.

Moreover, the cost of employing services like OpenAI highlights an important factor for broad adoption, especially when it comes to projects with limited resources. A possible obstacle to the accessibility of these technologies is the need for a comprehensive assessment of the costs and benefits of powerful artificial intelligence capabilities.

\subsection{Future work}

Building upon the insights gained from this research, our future work will explore several avenues:
\begin{enumerate}
    \item \textbf{Exploration of Open Source LLMs:} We plan to evaluate and integrate open-source LLMs to provide cost-effective alternatives and enable a comparative analysis of different models' performance. This will also enhance the system's natural language processing capabilities for improved accuracy and relevance in user story generation.
    
    \item \textbf{Incorporation of Diverse Prioritization Techniques:} To improve versatility and effectiveness, the tool will be expanded to support additional prioritization methodologies. This adaptation will cater to a broader range of user needs and research contexts.
    
    \item \textbf{User-Driven Story Generation:} The system will be further refined to generate user stories that precisely align with specific user requirements, ensuring relevance and actionability. Adaptive learning mechanisms will be incorporated to evolve the system based on user interactions and feedback.
     
    \item \textbf{Co-pilot Development for Requirement Generation and Prioritization:} A co-pilot feature will be built to automate the generation and prioritization of user requirements using advanced AI techniques such as RAG and lang-chain. This feature will provide users with a guided, interactive experience in defining and organizing their research activities.
     
\end{enumerate}

\section{Conclusions}
\label{conclusions}
In this study, we presented a tool that uses OpenAI, Flask, and React to automatically generate and rank user stories based on their fundamental requirements. The application marks a major advancement in using AI to improve project management workflows by allowing users to input requirements and then generate user stories and epics for prioritization through an AI agent. The system's ability to simplify project management activities is demonstrated by its ability to parse responses into an approachable JSON format and enable CSV downloads for connection with task management platforms like JIRA and Trello.


\begingroup
\setlength{\parskip}{0pt}
\setlength{\itemsep}{0pt plus 0.2ex}
\setlength{\bibsep}{0pt plus 0.3ex}
\bibliography{sn-bibliography}
\endgroup

\end{document}